\documentclass[conference]{IEEEtran}
\IEEEoverridecommandlockouts
\usepackage{cite}
\usepackage{amsmath,amssymb,amsfonts}
\usepackage{algorithmic}
\usepackage{graphicx}
\usepackage{textcomp}
\usepackage{xcolor}
\usepackage{color}

\def\BibTeX{{\rm B\kern-.05em{\sc i\kern-.025em b}\kern-.08em
    T\kern-.1667em\lower.7ex\hbox{E}\kern-.125emX}}
    
\IEEEoverridecommandlockouts
\IEEEpubid{\begin{minipage}{\textwidth}\ \\[80pt]
        \centering\normalsize{	© 2021 IEEE. Personal use of this material is permitted. Permission from IEEE must be
obtained for all other uses, in any current or future media, including
reprinting/republishing this material for advertising or promotional purposes, creating new
collective works, for resale or redistribution to servers or lists, or reuse of any copyrighted
component of this work in other works.}
\end{minipage}} 
\begin{document}

\title{Radio Environment Map and Deep Q-Learning for 5G Dynamic Point Blanking  \\

\thanks{The simulations were based on the QCM simulator from
Huawei Technologies Sweden Research Center.
The presented work has been funded by the Polish Ministry of Education and Science within the status activity task “Optimization of wireless network operation and compression of test data” in 2021, no. 0312/SBAD/8159.}
}

\author{\IEEEauthorblockN{ Marcin Hoffmann} 
\IEEEauthorblockA{\textit{Institute of Radiocommunications} \\
\textit{Poznań University of Technology}\\
Poznań, Poland \\
marcin.ro.hoffmann@doctorate.put.poznan.pl}
\and
\IEEEauthorblockN{ Paweł Kryszkiewicz}
\IEEEauthorblockA{\textit{Institute of Radiocommunications} \\
\textit{Poznań University of Technology}\\
Poznań, Poland \\
pawel.kryszkiewicz@put.poznan.pl}
}

\maketitle

\begin{abstract}
Dynamic Point Blanking (DPB) is one of the Coordinated MultiPoint (CoMP) techniques, where some Base Stations (BSs) can be temporarily muted, e.g., to improve the cell-edge users throughput. In this paper, it is proposed to obtain the muting pattern that improves cell-edge users throughput with the use of a Deep Q-Learning. The Deep Q-Learning agent is trained on location-dependent data. Simulation studies have shown that the proposed solution improves cell-edge user throughput by about 20.6\%. 
\end{abstract}


\section{Introduction}

The CoMP techniques are introduced for mobile networks in Release~11 of Long Term Evolution Advanced (LTE-A)~\cite{DAHLMAN201839}. The main idea behind the CoMP is that BSs cooperate, e.g., to improve the throughput of the cell edge users. Several CoMP techniques are investigated in the literature, e.g., Joint Transmission (JT) stands for simultaneous transmission to the User Equipments (UEs) from several BSs, or Dynamic Point Selection (DPS), where a UE is connected to a single BS but can be switched almost instantly to another BS~\cite{bassoy2017}. On the other hand, the much less investigated CoMP scheme is a so-called Dynamic Point Blanking (DPB). The idea of DPB is to mute particular BSs in frequency and time, e.g., to reduce interference, and improve the cell-edge users throughput~\cite{wang2015}. The challenge is to properly choose the muting pattern that is appropriate for given network conditions, e.g., the spatial distribution of users in a cell. This problem especially arises while considering the 5G Massive Multiple-Input Multiple-Output (M-MIMO) BSs in the Heterogeneous Network (HetNet). In such a network architecture inter-cell interference is much affected by the spatial channel correlations~\cite{sanguinetti2020}. 

The state-of-the-art solutions proposed for muting pattern selection in DPB mostly rely on the instantaneous Channel State Information (CSI) proper for single antenna BSs~\cite{wang2015}. This approach can be insufficient to model a realistic 5G M-MIMO HetNet consisting of several interacting functional blocks: precoder, user to BSs assignment, scheduler, and most importantly under a realistic, spatially-correlated radio channel. 

The aim of this paper is to propose an intelligent DPB algorithm, that targets the improvement of the cell-edge users' throughput. The proposed solution is based on the location-dependent data being stored and processed in the so-called Radio Environment Map (REM)~\cite{tengkvist2017}. Instead of using instantaneous CSI, the historical information about the observed cell-edge users' throughput related to the given spatial distribution of users, and the muting pattern is stored in REM. In our previous works, we have utilized the REM to implement a so-called table-based Reinforcement Learning scheme~\cite{hoffmann2021as}. In that approach, each spatial distribution of users, the muting pattern, and the resultant network performance metrics create a separate REM entry. However, while in a real network the number of possible user locations is large, the size of REM would rapidly grow. Thus in this work, we propose to train a so-called Deep Q-Network (DQN)~\cite{mnih2013playing}. The DQN still needs data from REM for the training purpose, but after the training phase, the DQN can potentially infer how to act under the unknown spatial distribution of users. The proposed solution based on DQN is validated within the realistic system-level simulator using an accurate 3D-Ray-Tracing radio channel model.

\section{Radio Environment Map and Deep Q-Learning} \label{sec:system_model}

\begin{figure}[htbp]
\centerline{\includegraphics[scale=0.4]{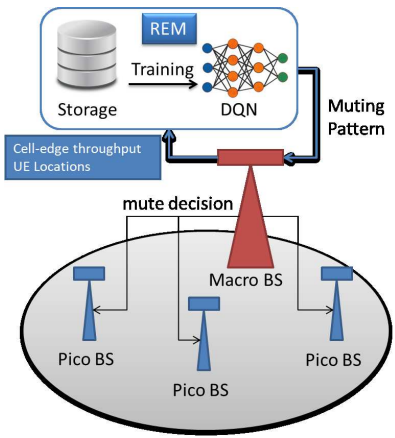}}
\caption{Dynamic Point Blanking based on REM and DQN.}
\label{fig:rem_concept}
\end{figure}
The diagram of the considered system is depicted in Fig.~\ref{fig:rem_concept}. It is assumed that the CoMP cluster covers a single HetNet cell consisting of one Macro BS (MBS) and several pico BSs (PBS). DPB process is centralized and managed by the MBS, which is not considered for muting to provide general coverage in a cell. Due to the assumption of ideal backhaul PBSs can be muted immediately. Just before muting a particular PBS, users connected to this PBS, are assigned to on of the active BSs, that provides the highest Received Signal Strength (RSS). The proposed algorithm aims to increase the cell-edge users' throughput (10th percentile from the distribution of user bitrates). To achieve this the MBS is extended with a dedicated REM module responsible for providing muting patterns proper for the currently observed spatial distribution of 
UEs. The REM consists of storage and the DQN model. During the learning phase, for each encountered spatial distribution of UEs, the REM tests all possible muting patterns and stores related cell-edge users' throughputs, i.e., an exhaustive search is performed. Later on, information from REM can be used to train a DQN. Training of DQN follows the standard RL scheme, where the agent observes the environment state, takes action, obtains a reward and the cycle repeats~\cite{sutton2018reinforcement}. The only modifications in relation to the standard tabular RL methods are that DQN takes the state as an input and outputs approximated rewards for all possible actions. The difference between the real reward and DQN-approximated rewards is used to update the weights of DQN. In this paper the RL scheme components are defined as follows:
\begin{itemize}
    \item \textbf{Environment} is the CoMP cluster in a single HetNet cell.
    \item \textbf{Agent} is a REM unit that chooses the muting pattern on the basis of users' location information. For this purpose Agent utilizes DQN. We assume that after the training phase agent acts greedy, i.e., takes action that will provide the best-approximated reward. 
    \item \textbf{State} can be defined directly as a raw set of UE locations. However, under such an approach state recognition is challenging for the DQN, due to the varying number of UEs in the cell, and the shuffle of UE coordinates. To overcome these issues we propose to pre-process the localization data with the use of a K-Means clustering algorithm~\cite{kanungo2002}. We split users into the number of clusters equal to the number of BSs in the considered HetNet cell. The algorithm starts with centers of clusters being set to the BSs' coordinates and outputs centers of clusters and the number of UE associated with each cluster. Flatten output of K-Means algorithm formulates state and is provided as an input to the DQN.      
    \item \textbf{Action} is defined as the combination of active PBSs. There are $2^{N}$ possible actions, where $N$ is the number of PBSs in the CoMP cluster.
    \item \textbf{Reward} is the throughput of the cell-edge users, defined as a 10-th percentile from the distribution of user bitrates. The reward is equal to 0 for actions that result in disconnecting of any UE from the network. This prevents such actions to be selected. 
\end{itemize}

\section{Simulation Results} \label{sec:results}

\begin{figure}[htbp]
\centerline{\includegraphics[height=5.5cm, width=8cm]{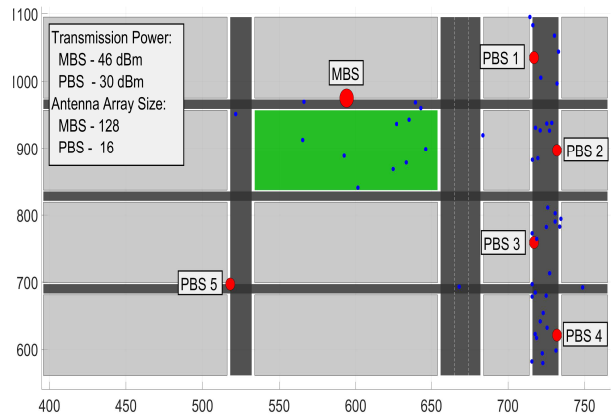}}
\caption{Deployment of BSs (red dots), and representative example of UE spatial distribution (blue dots).}
\label{fig:scenario}
\end{figure}

The proposed DPB algorithm based on REM and DQN is validated through computer simulations. The deployment of BSs and a representative example of UE spatial distribution are depicted in Fig.~\ref{fig:scenario}. It is assumed that a HetNet is operating in a common frequency band of 300 MHz at a carrier frequency of 3.5 GHz. There is one MBS equipped with a 128-antenna array, and 5 PBSs equipped with 16-antenna arrays. The MBS and PBSs have a transmission power of $46$~dBm, and $30$~dBm, respectively. We have assumed a \emph{full-buffer} traffic model, user to BSs association based on RSS, Regularized Zero-Forcing (RZF) precoder, and user scheduler aimed at providing proportional fairness. Radio channel coefficients are generated with the use of an accurate 3D-Ray-tracing model. More details on the simulation environment can be found in~\cite{HOFFMANN2021}.  

A total of 150 spatial distributions of UEs have been considered. Each exploits 50 UEs randomly distributed over the HetNet area (see Fig.~\ref{fig:scenario}). To provide REM with information about the cell-edge throughput related to all possible muting patterns, the simulation has been repeated 32 times under the same 150 UE spatial distributions. After that, the gathered data was used to train the DQN over 50000 epochs with a batch size of 8 samples. The DQN is designed arbitrarily to have four hidden layers of size 16, 64, 128, and 64, respectively.

\begin{figure}[htbp]
\centerline{\includegraphics[height=4.5cm, width=8cm]{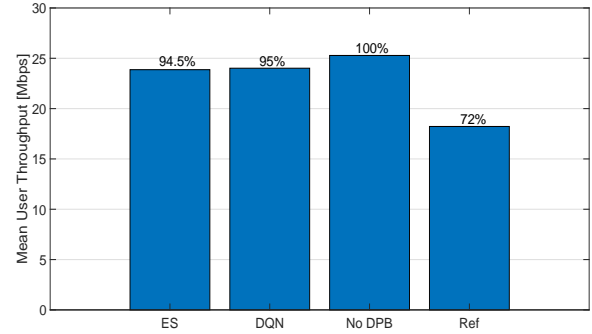}}
\caption{Mean user throughput averaged over 150 sets of randomly placed~UEs.}
\label{fig:mean_results}
\end{figure}

 Four algorithms have been evaluated. \emph{No DPB} refers to the case where DPB is not applied. \emph{Ref} stands for the DPB based on instantaneous CSI defined in~\cite{wang2015}. \emph{ES} refers to the utilization of an optimal muting pattern obtained on the basis of an exhaustive search. Finally, \emph{DQN} refers to the DPB scheme based on the pre-trained DQN that is stored in REM, and used for inference of the muting pattern. 
\begin{figure}[htbp]
\centerline{\includegraphics[height=4.5cm, width=9cm]{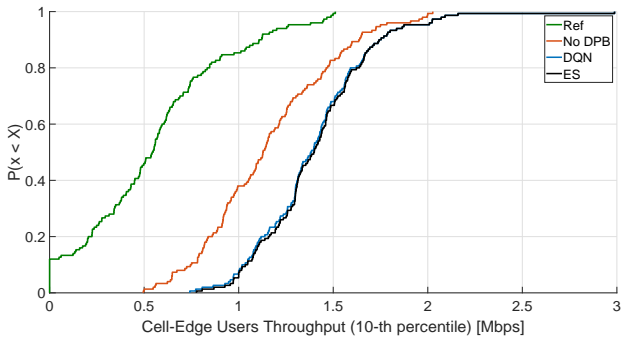}}
\caption{The Cumulative Distribution function of cell-edge users throughput observed over 150 sets of randomly placed UEs.}
\label{fig:results}
\end{figure}
\begin{figure}[htbp]
\centerline{\includegraphics[height=4.5cm, width=8cm]{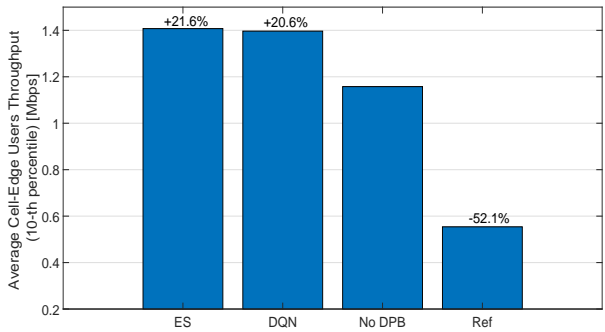}}
\caption{Cell-edge users throughput averaged over 150 sets of randomly placed~UEs.}
\label{fig:avg_results}
\end{figure}
In Fig.~\ref{fig:results}, there is a Cumulative Distribution function of cell-edge users throughput observed over 150 sets of randomly placed UEs for every algorithm. The observed states (UE locations) are the same, that were used during the training of DQN. It can be seen that \emph{Ref} algorithm has the worst performance because the used CSI is not accurate enough in approximating user throughput in the M-MIMO network. On the other hand, it is clearly visible that the proposed \emph{DQN} approach improved the cell-edge users throughput in every case. Moreover, the \emph{DQN} performs almost as well as the optimal \emph{ES} approach. Cell-edge throughput averaged over 150 sets of randomly placed UEs is presented in Fig.~\ref{fig:avg_results}. The \emph{Ref} algorithm due to its inability of accurate estimation of users throughput chooses the wrong muting patterns. This is why it causes about $52.1$\% degradation in cell-edge users throughput while comparing it against the \emph{No DPB} algorithm. On the other hand, the utilization of \emph{DQN} approach provides about $20.6$\% gain over the \emph{No DPB} algorithm. It is because the \emph{DQN} is trained on the historical data from the REM storage unit, that allows us to accurately infer the behavior of the system under various muting patterns. The \emph{DQN} achieves on average about $99$\% of cell-edge users throughput gains obtained under a selection of the optimal muting patterns provided by the \emph{ES} algorithm. 
Finally, in Fig.~\ref{fig:mean_results} there is presented a mean user throughput averaged over 150 sets of randomly placed UEs. It can be seen that in the case of a \emph{DQN} algorithm the cell-edge user throughput gains, comes with a cost of only 5\% degradation in mean user throughput while comparing to the scenario without DPB. On the other hand, the \emph{Ref} algorithm decreases the mean user throughput by about 28\%. 

\section{Conclusions} \label{sec:conclusions}

This paper shows that with the use of location-dependent information about cell-edge user throughput related to given muting patterns and UE spatial distributions stored in REM it is possible to effectively train and use DQN for the purpose of DPB. Simulations have shown that the proposed algorithm can improve the cell-edge users throughput by about $20.6$\% in relation to the algorithm without DPB. Moreover, the muting patterns selected by the DQN achieved 99\% of the exhaustive search performance. In the future, more extensive training of DQN will be conducted considering, e.g., various UE mobility patterns, and different configurations of BSs' antenna arrays.   

\bibliography{IEEEexample} 
\bibliographystyle{IEEEtran}

\end{document}